# Betti Number for Point Sets

Hao Wang
Ratidar Technologies LLC
* Corresponding author : haow85@live.com

**ABSTRACT**

Topology is the foundation for many industrial applications ranging from CAD to simulation analysis. Computational topology mostly focuses on structured data such as mesh, however unstructured dataset such as point set remains a virgin land for topology scientists. The significance of point-based topology can never be overemphasized, especially in the area of reverse engineering, geometric modeling and algorithmic analysis. In this paper, we propose a novel approach to compute the Betti number for point set data and illustrate its usefulness in real world examples. To the best of our knowledge, our work is pioneering and first of its kind in the fields of computational topology.

## 1. INTRODUCTION

Topology is one of the most successful math fields for the last hundreds of years. Better understanding of topology enables mankind to explore higher dimensional phenomena and advanced physics. Since the beginning of the IT revolution, computers are utilized to materialize topological ideas on machines. Computational topology are widely used in the fields of simulation analysis, computer assisted design, etc. However, due to the complexity of the theory, the technology is grasped only in a small group of experts compared with other technologies such as artificial intelligence. Most computational topology algorithms are confined to structured datasets such as mesh-based scalar fields or vector fields. Unstructured datasets such as point clouds are mostly neglected by the research community.

Point clouds are important because they exist as the primitive form of data acquired by 3D scanners and radars. Autonomous driving cars, BIM, 3D printing, among a whole range of industrial applications all rely heavily on effective processing and modeling of point clouds. A new research field named point-based graphics emerged in the 1990's and has flourished for more than a decade since then. However, point-based graphics basically ignores the important field of topology and emphasize other aspects such as non-topological modeling, rendering, etc. To the best of our knowledge, there has been no significant research paper on important point-based topological topics such as the computation of Betti numbers, and it is great regret for the entire scientific community. The latest development of computation topology mainly focus on research of topological quantities and structures on structured dataset, such as Morse-Smale Complex on scalar fields, and topological concepts on vector fields.

The major difficulty of creating point-based topological theories is the difficulty to quantify the mathematical quantities. The existence of points is usually correlated with the existence of noise, and therefore it introduces uncertainty into the formulation of the problem. In spite of all the problems, we make the first stride in the field in this paper to introduce a novel approach to compute the Betti number of a point cloud. After overview of our mathematical formulation, we demonstrate the application of our new idea in real world examples.

## 2. RELATED WORK

Computational topology is a field developed for decades. Researchers such as Pascucci [1][2][3] and Edelsbrunner [4][5] has developed computational Morse Theory as well as vector field topological analysis approaches that are widely applied in the field of manufacturing and simulation analysis. Although basic concepts of topology such as Betti number and Morse-Smale complex have been well known, relevant work on point data is basically neglected by the research community.

Point data algorithms are mostly research work in computer graphics [6][7] and computer vision [8][9][10]. Pioneering work in the field includes Moving Least Squares [11][12][13], etc. Point data algorithms are widely adopted in the application scenarios of reverse engineering and autonomous driving car industry.

In the field of financial engineering , extreme value theory [14][15] is a sub-field aiming to capture the extreme events in the financial industry. The idea could be easily carried over to the machine learning, as power law effect exists almost everywhere in the real world. In this paper, we use Moment Estimator [16] , QQ Estimator [17] and Peng Estimator [18] in our experiments.

## 3. BETTI NUMBER

The most straight-forward explanation of Betti number is the number of holes in a manifold. For example, the Betti number of a sphere is 1. In our approach we first compute the radius of the circle inscribed in each hole candidate, and then compute the Degree of Matthew Effect of the radius list. We use the computed tail index estimator as a statistical approximation for the Betti number.

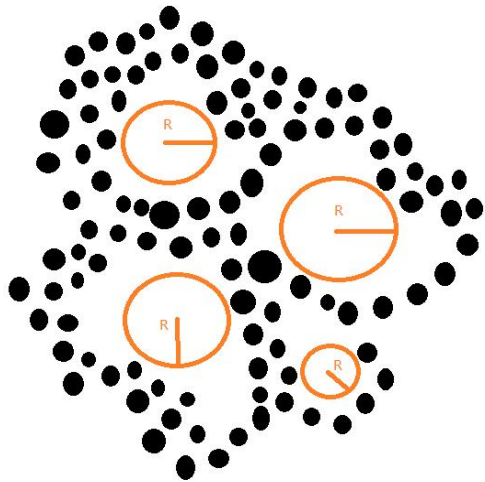

Fig. 1 A Point Set of Betti Number 4

Fig. 1 shows a point set of Betti number 4. We compute the distance between every pair of points in the dataset and take its one half and use it as the candidate for a potential hole's radius. As illustrated in Fig.1 , each of the 4 holes' radii is included in the candidate pool. We rank the half-distances by its values and then borrow the idea of Extreme Value Theory from financial engineering to compute the statistical estimate for Betti number. Since the number of holes is small compared with the number of points in the point cloud and also the radii of holes are much larger than other point pairs. This phenomenon is in consistent with the power law effect, and this is why we choose the tail index estimators as our estimator for Betti number. Extreme Value Theory quantifies the power of exponent of the power law distribution. It is an accurate measure of the skewness of the power law distribution.

We illustrate our idea of computing Betti number on several classic manifold example in the following section, and then demonstrate the application of the idea in the real world examples.

## 4. MANIFOLD EXAMPLES

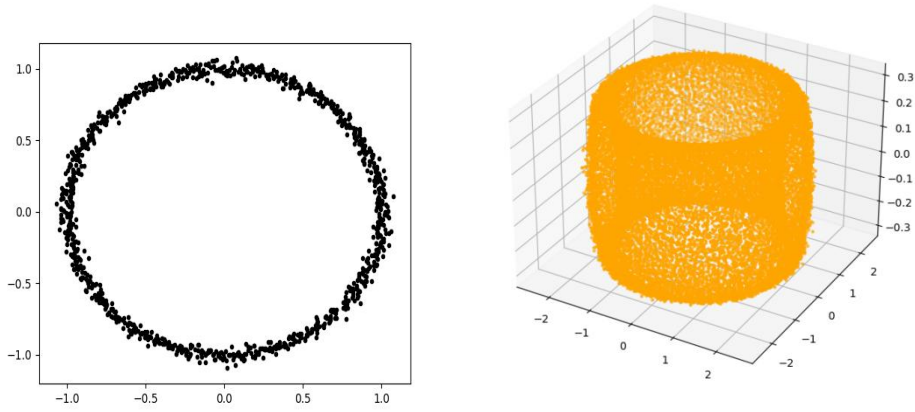

Fig. 2 Two manifold examples to demonstrate the idea of Betti number for point clouds - A 2D circle and a 3D torus.

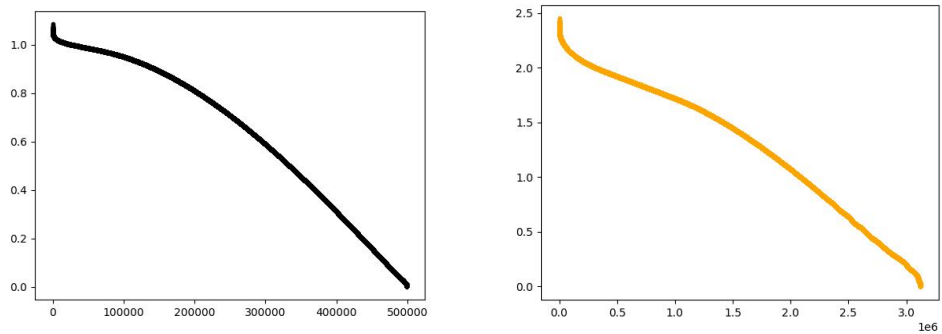

Fig. 3 Distribution of half values of pair-wise distances.

Fig. 2 shows 2 examples of point-based manifolds - a 2D circle with noise and a 3D torus with noise, and Fig. 3 demonstrates distributions of the sorted radii list of the 2 manifolds. We compute the extreme value estimators for the 2 curves, and obtain the following results :

|  | Moment Estimator | QQ Estimator | Peng Estimator |
|---|---|---|---|
| Circle | -0.62650 | 0.00140 | 1.00146 |
| Torus | -0.28842 | 0.00195 | 1.00195 |

Table 1 Extremal Value Estimators for Circle and Torus

From Table 1, we observe that 3D torus has a Betti Number estimator larger than a 2D circle. In our example, the ground-truth Betti Number for 3D torus should be 2, while the ground-truth Betti Number for 2D circle is 1. Although our extreme value estimator for Betti Number is not precise, it serves as a comparative metric.

## 5. REAL WORLD DATASETS

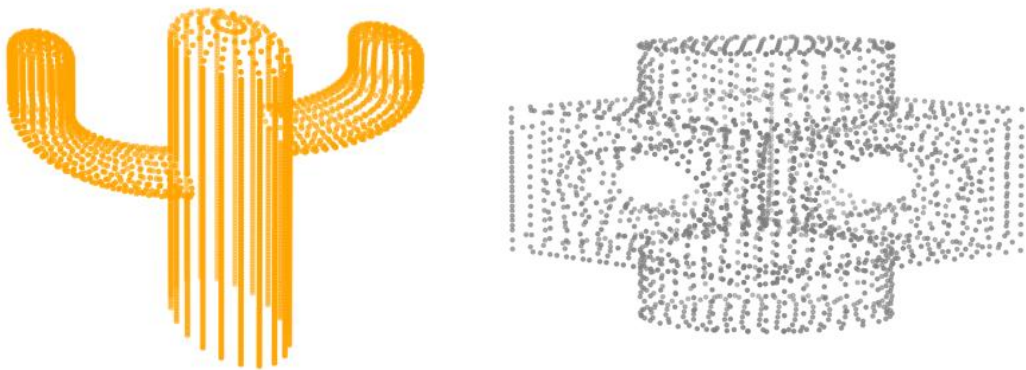

Fig. 4 Real World Examples of Point Sets

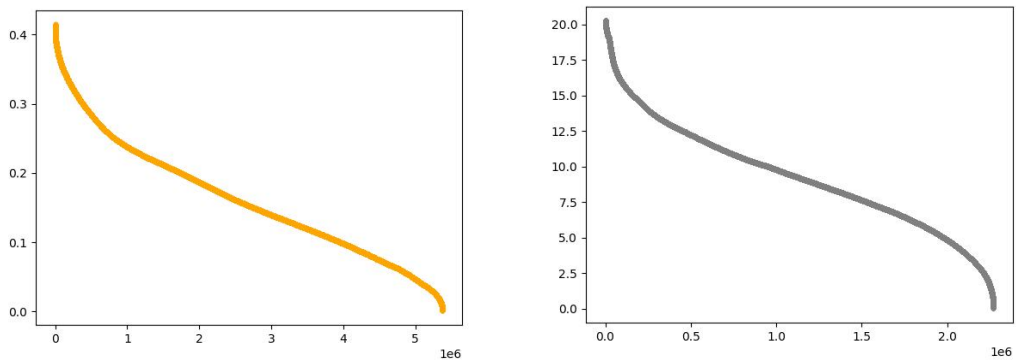

Fig. 5 Real World Examples of Point Sets

We test our novel idea on the real world datasets as demonstrated in Fig. 4 and plot the halves of pairwise distances in Fig. 5. We show the results of extremal value estimators in Table 2 :

|  | Moment Estimator | QQ Estimator | Peng Estimator |
|---|---|---|---|
| Cactus | -0.34205 | 0.00124 | 1.00238 |
| Tool | -0.77971 | 0.00054 | 1.00083 |

Table 2 Betti Number for Real World Examples

The experimental results tell us Tool has a smaller Betti Number than Cactus, which is consistent with our observation of the point clouds. We have shown that our approach is workable not only for synthesized examples, but also real world datasets.

## 6. DISCUSSION

From the experiments, we notice that when the Betti Number for ground truth is larger, the Betti Number estimator becomes larger. The Betti Number estimator itself is not a precise indicator of the ground truth, but it is a comparative metric for comparing, ranking and ordering tasks.

There are many different tail index estimators in the theory of extreme value, but not everyone of them is stable when used in our approach. We suggest the readers to check more than 5 different estimators and choose the majority votes as the final result. Famous estimators other than the ones mentioned in this paper include Pickands Estimator, Hill Estimator, and Adapted Hill Estimator. Actually the experimental results in this paper are already selected majority result among a candidate pool of estimators.

Our algorithm could also be applied to algorithmic analysis. For example, we can apply Takens Embedding [19] to MAE / RMSE curve of recommender systems, so the results would be elevated into 2D or 3D spaces, in which we could compute the Betti Number estimator for the point cloud, so we would be able to know the topological complexity of the algorithmic output.

We notice that the presence of noise serves as a major trouble just like the discreteness of the datasets. However, our approach is resilient to the existence of noise in the datasets, as shown in the previous 2 sections.

## 7. CONCLUSION

In this paper, we discuss a new idea of computing Betti Number estimator for point clouds. We borrow the idea from the field of financial engineering to facilitate the computation of our metrics, and show in experiments that our novel approach is effective.

In future work, we would like to explore the methods to compute other topological quantities and structures on point data such as Morse-Smale complex. We would also like to investigate into the uncertainty issue of the problem.